\documentclass[twocolumn,showpacs,amsmath,amssymb,floats,floatfix,prb,aps,superscriptaddress,showkeys]{revtex4}

\usepackage{graphicx}
\usepackage{epsfig}
\usepackage{dcolumn}
\usepackage{bm}
\usepackage{times}
\graphicspath{{C:/Users/Gabriel/Desktop/figures/}}
\DeclareMathOperator{\sech}{sech}
\begin{document}
\title{Dirac equation in one dimensional transformation optics}
\author{Gabriel Gonz\'alez}\email{gabriel.gonzalez@uaslp.mx}
\affiliation{C\'atedras CONACYT, Universidad Aut\'onoma de San Luis Potos\'i, San Luis Potos\'i, 78000 MEXICO}
\affiliation{Coordinaci\'on para la Innovaci\'on y la Aplicaci\'on de la Ciencia y la Tecnolog\'ia, Universidad Aut\'onoma de San Luis Potos\'i,San Luis Potos\'i, 78000 MEXICO}
\begin{abstract}
We show that the propagation of transverse electric (TE) polarized waves in one dimensional inhomogeneous settings can be written in the form of the Dirac equation in one space dimension with a Lorentz scalar potential, and consequently perform photonic simulations of the Dirac equation in optical structures. In particular, we propose how the zero energy state of the Jackiw-Rebbi model can be implemented in a optical set up by controlling the refractive index landscape, where TE polarized waves mimic the Dirac particles and the soliton field can be implemented and tuned by adjusting the refractive index.
\end{abstract}

\maketitle

\begin{figure}
\centering
\includegraphics[width=\linewidth,height=4cm]{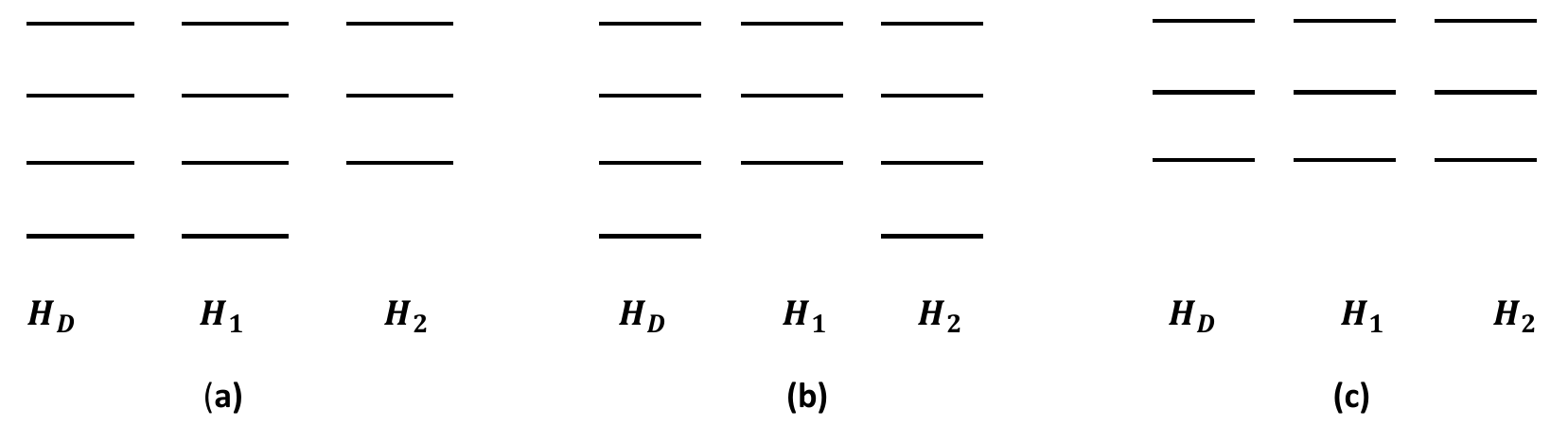} 
\caption{Schematic diagram of the possible allignment of energy spectra of the Hamiltonians $H_D$, $H_1$ and $H_2$ are shown when (a) there exist a zero energy mode and $\psi_1$ is normalizable, (b) there exist a zero energy mode and $\psi_2$ is normalizable and (c) when there is no zero energy mode and neither $\psi_1$ nor $\psi_2$ is normalizable.}
\label{fig1}
\end{figure}

The Dirac equation is one of the fundamental equations in theoretical physics that accounts fully for special relativity in the context of quantum mechanics for elementary spin-1/2 particles.\cite{pam} The Dirac equation plays a key role to many exotic physical phenomena such as graphene,\cite{novo} topological insulators\cite{topo} and superconductors.\cite{topos} These systems proved to be ideal testing grounds for theories of the coexistence of quantum and relativistic effects in condensed matter physics.\\
More recently, with the advances of experimental and material science techniques, a collection of effects in different fields have been simulated using different physical platforms such as optical structures,\cite{ucf1} metamaterials\cite{wei} and ion traps.\cite{lama} \\
The purpose of this letter is to demonstrate that optics can provide a fertile ground where physical phenomena described by the Dirac equation can be explored. In particular, we demonstrate that the TE polarized electromagnetic waves in one dimensional inhomogeneous media can be mapped into the Dirac equation in one dimension with a Lorentz scalar potential. By tailoring the refractive index we propose a optical structure that simulates a historically important relativistic model known as the Jackiw-Rebbi model.\cite{jackreb} The model describes a one dimensional Dirac field coupled to a static background soliton field and is known as one of the earliest theoretical description of a topological insulator where the zero energy mode can be understood as the edge state. The Jackiw-Rebbi model can be equivalently thought of as the model describing a massless Dirac particle under a Lorentz scalar potential. In particular, the Jackiw-Rebbi model has been studied by Su, Shrieffer and Heeger in the continuum limit of polyacetylene.\cite{shrieffer}\\
To explore the connection between the Dirac equation and optical wave propagation in one dimension with an arbitrary refractive index distribution $n(x)$ we consider TE waves propagating in the $xz$ plane. Field modes propagating in this system are described by the following Helmholtz equation\cite{em}
\begin{equation}
(\partial_{xx}+\partial_{zz}+k_0^2n(x))E_y(x,z)=0,
\label{eq01}
\end{equation} 
where $k_0$ is the vacuum wavenumber. TE modes governed by eq.(\ref{eq01}) have the form $E_y(x,z)=\psi_1(x)e^{i\beta z}$, where $\beta=k_0n_0\sin\theta$ is the propagation constant, $n_0$ is the constant background value of the refractive index at $x\rightarrow\pm\infty$, $\theta$ is the angle of incidence, and $\psi_1$ satisfy the following Schrodinger like equation\cite{ucf2}
\begin{equation}
\left(\frac{\hat{k}^2}{k_0}+U_1(x)\right)\psi_1(x)=0
\label{eq02}
\end{equation}
where $\hat{k}=i\partial_x$, $U_1(x)=-k_0(n^2(x)-n^2_0\sin^2\theta)$. Let us now make the following transformation
\begin{equation}
U_1(x)=\frac{1}{k_0}\left[\left(\frac{k_0}{2}+S(x)\right)^2-E^2-\frac{dS}{dx}\right]
\label{eq03}
\end{equation}
where $S(x)$ is a Lorentz scalar function and $E$ is an auxiliary constant. Substituting eq.(\ref{eq03}) into eq.(\ref{eq02}) we have
\begin{equation}
\psi_1^{\prime\prime}-\left(k_0S+S^2-S^{\prime}\right)\psi_1+\left[E^2-\left(\frac{k_0}{2}\right)^2\right]\psi_1=0
\label{eq04}
\end{equation}
Adding and subtracting the term $(k_0/2+S)\psi_1^{\prime}$ to the left hand side of eq.(\ref{eq04}) we have
\begin{equation}
\left[\psi_1^{\prime}+(k_0/2+S)\psi_1\right]^{\prime}-(k_0/2+S)\psi_1^{\prime}+\left[E^2-(k_0/2+S)^2\right]\psi_1=0
\label{eq05}
\end{equation}
If we make the following substitution $\psi_1^{\prime}+(k_0/2+S)\psi_1=E\psi_2$ into eq.(\ref{eq05}) we end up with the following equation $-\psi_2^{\prime}+(k_0/2+S)\psi_2=E\psi_1$. These two coupled differential equations can be written in the same mathematical form as the Dirac equation with $c=\hbar=1$, i.e.
\begin{equation}
\hat{H}_D\Psi=\left[\sigma_y \hat{p}+\sigma_x\left(\frac{k_0}{2}+S(x)\right)\right]\Psi=E\Psi
\label{eq06}
\end{equation}
where
\begin{eqnarray} 
\sigma_y=\left(\begin{array}{cc} 0 & -i\\ i & 0 \end{array}\right),\quad \sigma_x=\left(\begin{array}{cc} 0 & 1\\ 1 & 0 \end{array}\right) \quad \mbox{and} \quad \Psi=\left(\begin{array}{c}\psi_1\\ \psi_2\end{array}\right).
\label{eq07}
\end{eqnarray}
Equations (\ref{eq06}) can be reduced to two uncoupled Schrodinger equations $\hat{H}_{i}\psi_{i}=0$, for $i=1,2$, given by
\begin{equation}
\hat{H}_i\psi_i=\left(\frac{\hat{k}^2}{k_0}+U_i(x)\right)\psi_i(x)=0
\label{eq08}
\end{equation}
where
\begin{equation}
U_2(x)=\frac{1}{k_0}\left[\left(\frac{k_0}{2}+S(x)\right)^2-E^2+\frac{dS}{dx}\right].
\label{eq09}
\end{equation}
Clearly, $\hat{H}_{1,2}$ are supersymmetric partner Hamiltonians which can be factorized as $\hat{H}_1=\hat{A}^{\dagger}\hat{A}-(E^2/k_0)$ and $\hat{H}_2=\hat{A}\hat{A}^{\dagger}-(E^2/k_0)$ where $\hat{A}=(\partial_x+k_0/2+S(x))/\sqrt{k_0}$ and $\hat{A}^{\dagger}=(-\partial_x+k_0/2+S(x))/\sqrt{k_0}$. 
Thus, if $E_{1(2)}$ is an eigenvalue of $\hat{H}_1(\hat{H}_2)$ with eigenfunction $\psi_1(\psi_2)$, the same eigenvalue is given for $\hat{H}_2(\hat{H}_1)$ with corresponding eigenfunction $\hat{A}\psi_1(\hat{A}^{\dagger}\psi_2)$. The only exception is the ground state of $\hat{H}_1$ which lacks a counterpart in $\hat{H_2}$. \\
From eq.(\ref{eq06}) it follows that the Hamiltonian $H_D$ possesses a {\it chiral} symmetry defined by the operator $\sigma_z$, which anticommutes with the Hamiltonian, i.e. $\left\{\hat{H}_D,\sigma_z\right\}=0$. The {\it chiral} symmetry implies that eigenstates come in pairs with $\pm E$. It is possible however for an eigenstate to be its own partner for $E=0$, if this is the case then the state is topologically protected. The resulting zero energy state is protected by the topology of the scalar field, whose existence is guaranteed by the index theorem, which is localised around the soliton.\cite{jackreb}\\
We can easily construct the zero energy mode by setting $E=0$ in eq.(\ref{eq06}) and solving for the uncoupled first order differential equations for $\psi_{1,2}$, i.e.
\begin{equation}
\psi_{i}=C_{\mp}\exp\left[\mp\int\left(\frac{k_0}{2}+S(x)\right)dx\right]
\label{eq10}
\end{equation}
where $C_{\mp}$ is a normalization constant and the double sign in eq.(\ref{eq10}) is $-(+)$ for $i=1(2)$. Note that $\psi_{1,2}$ cannot be both normalized. In the case when neither $\psi_1$ nor $\psi_2$ is normalizable there are no zero modes allowed, and $\hat{H}_{1,2}$ share the same energy spectrum. In the case when $\psi_1$ is normalizable and $\psi_2$ is not, then there is a zero mode allowed and $\hat{H}_{1,2}$ share the same energy spectrum except for the ground state of $\hat{H}_1$. In Fig.(\ref{fig1}) we show the energy spectra when there is or there is not a zero energy state for the Dirac Hamiltonian.\\ The existence of a zero mode then depends on the asymptotic behavior of $(k_0/2+S(x))$, in general we have that\cite{susyd}
\begin{equation}
\frac{k_0/2+S(\infty)}{k_0/2+S(-\infty)}=\left\{\begin{array}{ll}
																									+1, \mbox{there is no zero mode},\\
																									-1, \mbox{there exists a zero mode}.
																									\end{array}\right.
\label{eq11}
\end{equation}
We are interested in the zero energy mode of the Jackiw-Rebbi model which is described by the following Dirac equation\cite{jackreb}
\begin{equation}
\left(\begin{array}{cc}
\partial_x+\phi(x) & 0\\ 0 & -\partial_x+\phi(x)\end{array}\right)\left(\begin{array}{c}\psi_1(x) \\ \psi_2(x)\end{array}\right)=0 
\label{eq12}
\end{equation} 
where $\phi(x)=m\tanh(\lambda x)$ corresponds to the soliton localised at $x=0$, with $m>0$ and $\lambda>0$. If we take the ``superpotential" $k_0/2+S(x)=\phi(x)$ then we see from eq.(\ref{eq11}) that there exists a zero mode with the following solutions
\begin{equation}
\psi_1(x)=C_-\left[\cosh(\lambda x)\right]^{-m/\lambda} \, \mbox{and} \, \psi_2(x)=C_+\left[\cosh(\lambda x)\right]^{m/\lambda}. 
\label{eq13}
\end{equation}
We need to set $C_+=0$ in order to make the two-component spinor normalizable. Therefore, the wave function for the zero mode is given by
\begin{equation}
\Psi(x)=C_-\left(\begin{array}{c} \left[\cosh(\lambda x)\right]^{-m/\lambda} \\ 0\end{array}\right).
\label{eq14}
\end{equation}
Substituting the ``superpotential" into eq.(\ref{eq03}) and using the fact that $U_1(x)=-k_0(n^2(x)-n^2_0\sin^2\theta)$ we can get the expression for the refractive index, i.e.
\begin{equation}
n^2(x)=\left(\frac{m\lambda}{k_0^2}+\frac{m^2}{k_0^2}\right)\sech^2(\lambda x)-\frac{m^2}{k_0^2}\sec^2\theta.
\label{eq15}
\end{equation}
\begin{figure}[ht]
    \centering
        \includegraphics[width=\linewidth,height=7cm]{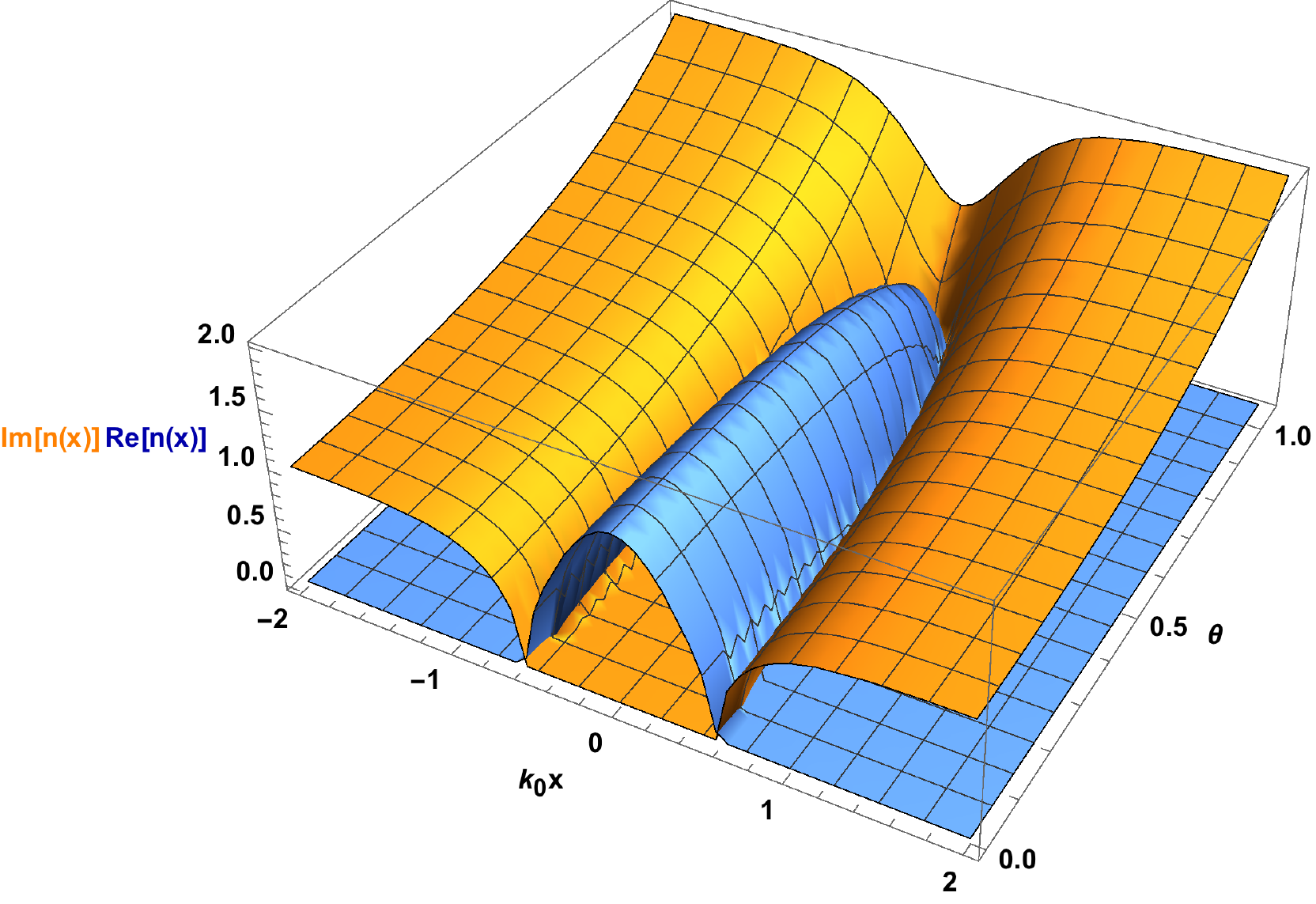} 
		\caption{Real and Imaginary parts of the refractive index profile obtained from the Jackiw-Rebbi model as a function of $k_0x$ and the incidence angle $\theta$. Note how the real and imaginary part are both even functions with respect to the $x$ coordinate. We have used $\lambda=2m$ and $m=k_0$.}
\label{fig2}
\end{figure}
In Fig.(\ref{fig2}) we show the real and imaginary parts of the refractive index obtained from the Jackiw-Rebbi model as a function of the coordinate and angle of incidence.
From eq.(\ref{eq15}) we see that $n_0=im\sec\theta/k_0$, which means that the TE mode propagating in a optical structure with a refractive index given by eq.(\ref{eq15}) which mimicks the zero-mode state of the Jackiw-Rebbi model is an evanescent wave of the form $E_y(x,z)=\left[\cosh(\lambda x)\right]^{-m/\lambda}e^{-m\tan\theta z}$ (See Fig.(\ref{fig3})). Note that if we have $m<0$ the results remain exactly the same except that we must set $C_-=0$ in order to make the two-component spinor normalizable.\\
\begin{figure}[b]
    \centering
        \includegraphics[width=\linewidth,height=7cm]{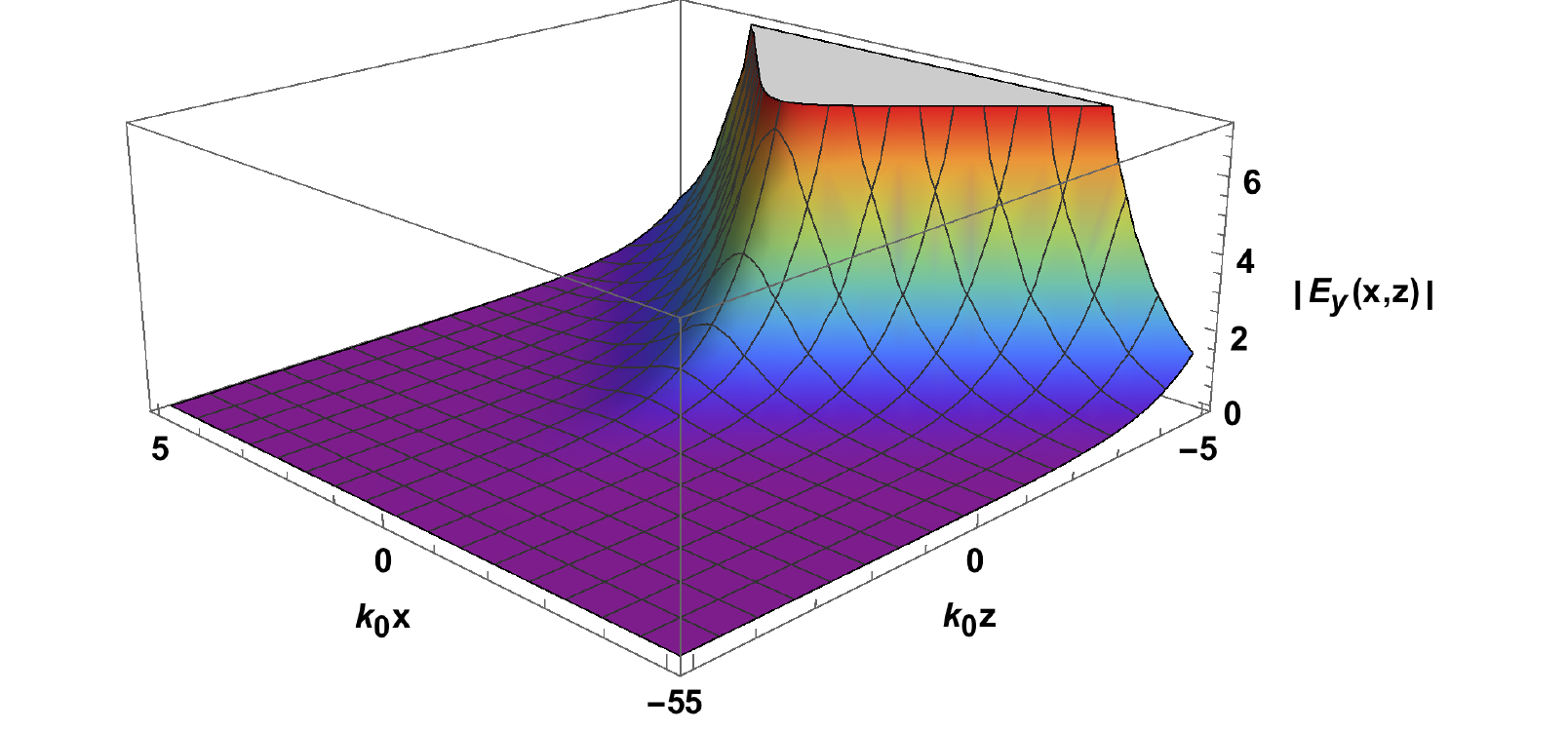} 
		\caption{Intensity evolution of the TE polarized wave inside a waveguide with a refractive index which mimics the Jackiw-Rebbi model. We have set $\lambda=2m$, $m=k_0$ and $\theta=\pi/4$.}
\label{fig3}
\end{figure}
It is well known that the refractive index is in general a complex function and this fact invites us to explore if there is a zero-energy state for a given complex ``superpotential".
Let us express then a complex ``superpotential" of the form $k_0/2+S(x)=a(x)+ib(x)$, where $a(x)$ and $b(x)$ are real valued functions. Then, we can have two different ``optical" potentials given by\cite{ptsym}
\begin{equation}
U_{1,2}(x)=\frac{1}{k_0}\left[\left(a^2-b^2\mp \frac{da}{dx}\right)+i\left(2ab\mp\frac{db}{dx}\right)\right].
\label{eq16}
\end{equation} 
By employing a complex ``superpotential" we are entering into the field of Non-Hermitian Hamiltonians, fortunately a consistent quantum theory can be constructed for Non-Hermitian Hamiltonians which possess parity-time ($\mathcal{PT}$) symmetry.\cite{bender} The effect of $\mathcal{P}$ is to make spatial reflections, i.e. $p\rightarrow-p$ and $x\rightarrow-x$, and the effect of $\mathcal{T}$ is to perform complex conjugation. Hence, $\mathcal{PT}$ invariance of the Hamiltonian $H_{1,2}$, i.e. $[\mathcal{PT},H_{1,2}]=0$, requires $U_{1,2}(-x)=U^{*}(x)$. 
We propose the following two functions for $a(x)=-m\tanh(2m x)$ and $b(x)=2m\sech(2m x)$ which fulfills eq.(\ref{eq11}) and guarantees that a zero energy mode exists. Substituting $a(x)$ and $b(x)$ into eq.(\ref{eq16}) we have
\begin{eqnarray}
U_1(x)\!\!&=&\!\!-\frac{m^2}{k_0}-\frac{3m^2}{k_0}\sech^2(2m x), \\ 
U_2(x)\!\!&=&\!\!U_1-i\frac{8m^2}{k_0}\sech(2m x)\tanh(2m x).
\label{eq17}
\end{eqnarray}
Note that we have chosen $a(x)=b^{\prime}/2b$ in order for $U_1(x)$ to be a real function and $U_{1,2}(x)$ to be $\mathcal{PT}$-symmetric. Using eq.(\ref{eq10}) it is clear that the normalizable zero energy state which corresponds to $U_2(x)$ is given by
\begin{equation}
\psi_2(x)=C_+\sqrt{\sech(2m x)}e^{i\varphi(x)}
\label{eq18}
\end{equation}
where $\varphi(x)=2\tan^{-1}(\tanh(mx))$ is a phase factor and $C_+$ is a normalization constant. Using the fact that $U_2(x)=-k_0(n^2(x)-n^2_0\sin^2\theta)$ we can get the expression for the complex refractive index, i.e. 
\begin{eqnarray}
\label{eq19}
n^2(x)&=&\frac{3m^2}{k_0^2}\sech^2(2m x)-\frac{m^2}{k_0^2}\sec^2\theta+\\ \nonumber
 & &  i\frac{8m^2}{k_0^2}\sech(2mx)\tanh(2mx).
\end{eqnarray}
In Fig.(\ref{fig4}) we show the real and imaginary parts of the refractive index given by eq.(\ref{eq19}).
In particular, we see that both refractive indices given by eq.(\ref{eq15}) and eq.(\ref{eq19}) give the same normalizable zero energy state, for the case when $\lambda=2m$, up to a phase factor. Therefore, both refractive indices will give us the same electric field norm that mimicks the Jackiw-Rebbi model. Interestingly, the refractive index given by eq.(\ref{eq19}) is complex $\mathcal{PT}$ symmetric. Recently, the physical realization of complex $\mathcal{PT}$ symmetric periodic potentials were investigated within the context of optics, which makes this proposal accessible for detecting topological states in the optical domain.\cite{ucf3,ucf4}\\
\begin{figure}
    \centering
        \includegraphics[width=\linewidth,height=7cm]{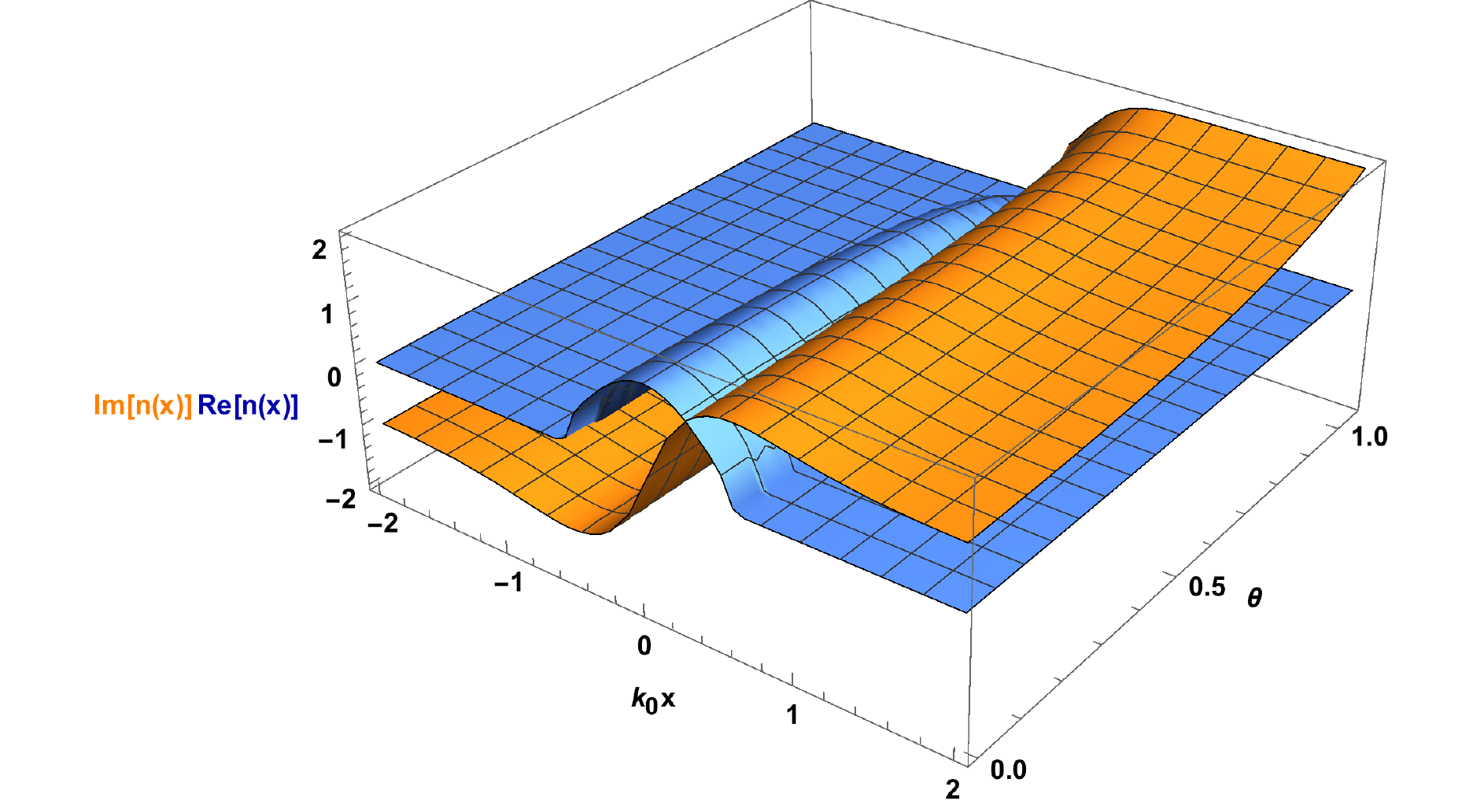} 
		\caption{Real and Imaginary parts of the refractive index profile obtained from the complex Jackiw-Rebbi model as a function of $k_0x$ and the incidence angle $\theta$. Note how the real part is even and the imaginary part is odd with respect to the $x$ coordinate, which exhibits the $\mathcal{PT}$ symmetric invariance of the refractive index.}
\label{fig4}
\end{figure}
In conclusion we have shown that TE polarized waves in one dimensional inhomogeneous settings can be used to simulate the dynamics of the Dirac equation in one space dimension with a Lorentz scalar potential. In particular, we demonstrate how the zero energy state of the Jackiw-Rebbi model can be implemented in a designed optical set up with a specific refractive index. We have also shown that the zero energy state of the Jackiw-Rebbi model can be reproduced with a complex effective mass. Based on these findings, we have introduced an optical platform for engineering topological states in the optical domain by controlling the refractive index landscape, in particular we propose a way for directly realizing the Jackiw-Rebbi model which allows one to probe the topologically protected zero energy mode.
\section{Acknowledgments}
This work was supported by the program ``C\'atedras CONACYT". The author gratefully acknowledge useful discussions with Fco. Javier Gonz\'alez.

\end{document}